\documentclass[aps,prl,twocolumn,showpacs,superscriptaddress,groupedaddress]{revtex4-1}  

\usepackage{longtable}
\usepackage{morefloats}
\usepackage[dvips]{graphicx}
\usepackage{color}
\usepackage{epsfig,graphicx,amsfonts,amsbsy}
\usepackage{amsmath,amsfonts,amsthm,amssymb}
\usepackage{appendix}
\usepackage{makeidx}
\usepackage{url}
\usepackage{verbatim}
\usepackage[bookmarksnumbered,pdfpagelabels=true,plainpages=false,colorlinks=true,linkcolor=blue,citecolor=red,urlcolor=blue]{hyperref}
\usepackage[rightcaption]{sidecap}
\usepackage{array}
\usepackage{booktabs}
\usepackage{multirow}
\usepackage{tabularx}
\usepackage{cancel,soul,ulem}

\newcommand{\bs}[1]{\boldsymbol{#1}}

\makeatletter

\begin{document}
\title{\bf{Antiferromagnetic Single-layer Spin-Orbit Torque Oscillators}}
\author{Roberto E. Troncoso$^{1}$}
\author{Karsten Rode$^{2}$}
\author{Plamen Stamenov$^{2}$}
\author{J. Michael D. Coey$^{2}$}
\author{Arne Brataas$^{1}$}
\affiliation{$^{1}$Center for Quantum Spintronics, Department of Physics, Norwegian University of Science and Technology, NO-7491 Trondheim, Norway}
\email{r.troncoso@ntnu.no}
\affiliation{$^{2}$CRANN, AMBER and School of Physics, Trinity College Dublin, Dublin 2, Ireland}
	
\begin{abstract}
We show how a charge current through a single antiferromagnetic layer can excite and control self-oscillations. Sustained oscillations with tunable amplitudes and frequencies are possible in a variety of geometries using certain classes of non-centrosymmetric materials that exhibit finite dissipative spin-orbit torque. We compute the steady-state phase diagram as a function of the current and spin-orbit torque magnitude. The anisotropic magnetoresistance causes the conversion of the resulting AF oscillations to a terahertz AC output voltage. These findings provide an attractive and novel route to design terahertz antiferromagnetic spin-orbit torque oscillators in simple single-layer structures.
\end{abstract}
\maketitle

{\it Introduction.}-- Modern technologies cannot easily generate and detect radiation in the spectral range between radio waves and infrared light \cite{TH_roadmap}. Typically, microwaves reach at most 100 GHz while optical techniques cover frequencies higher than 30 THz \cite{Tonouchi2007}. Despite enormous efforts\cite{TH_roadmap}, in this `Terahertz gap', efficient, cheap, chip-sized and practical technologies do not exist \cite{Tonouchi2007}. 
	
Spin dynamics of antiferromagnets (AFs) have the potential to fill the Terahertz gap \cite{Nunez2006,Baltz2018rmp} as their spin excitation frequency lie in the required range \cite{TH_roadmap}. They are robust against external fields since there is no net magnetic moment \cite{Baltz2018rmp}, but on the other hand, this implies there is no simple way to control them by magnetic fields \cite{Antiferromagnets2016}. As a result, current-induced and -controlled AF spin dynamics has emerged \cite{Wolf2001,Antiferromagnets2016,Baltz2018rmp}. To demonstrate their potential as enabling components in terahertz devices, we must establish how to control ultra-fast switching and oscillations \cite{Ivanov2014,Satoh2010,Wienholdt2012,RCheng:2015prb}.
	
Single-layer spin-orbit torques offer such possibilities \cite{Hals2013,zelezny2014,zelezny2017,ci-sot-2018}. The origin of these torques is the relativistic spin-orbit interaction. In non-centrosymmetric AFs, where the magnetic sublattices form inversion partners, a current induces a spin polarization with an alternating sign between the inversion partner lattice sites. There are recent observations of field-like (reactive) SOTs, dubbed N\'eel spin-orbit torque (NSOT), in CuMnAs \cite{wadley2016} and Mn${}_2$Au \cite{Mn2Au2018}. The functional form of SOTs depends on the crystallographic symmetries \cite{Hals2013}, and we expect interesting phenomenology in a wide variety of AFs. 

In addition to reactive SOTs, the transfer of angular momentum from currents to magnetization can compensate or enhance the magnetic (Gilbert) damping via dissipative (`damping-like') torques.  This can result in steady-state magnetic oscillations, also known as self-oscillations \cite{JENKINS2013}. Spin-transfer torques have a large dissipative component and can therefore sustain stable, high-frequency oscillations \cite{RanCheng2016,Khymyn2017,Sulymenko2017,Jakub2017}, but generally require an external source of spin-currents. Radically different, single-layer AF-based SOT oscillators would not need a spin polarizer, and has the potential of combining stable oscillations with high speed dynamics, useful for emitters and receivers of THz radiation. 
	
In this Letter, we propose a new route to realize current-controlled SOT oscillators in AFs. First, we determine the form of the SOTs based on symmetry arguments through a phenomenological approach. Second, the behaviour of self-oscillations is studied for different ratios of the reactive and dissipative torques as well as its dependence on the charge current density.
	
{\it Spin-orbit torques.}-- We consider a two-sub-lattice, non-centrosymmetric AF described by unit vectors along the magnetic moments, ${{\bs m}_a}$ and ${{\bs m}_b}$. A general form of the current-induced torque \cite{Hals2013,ci-sot-2018} in the local approximation is ${\bs\tau}_{\alpha}={\bs m}_{\alpha}\times{\bf H}_{\alpha}\left[{\bs m}_{a},{\bs m}_{b},{\bf j}\right]$, where $\alpha$ labels the sublattices. ${\bf H}_{\alpha}$ is the effective field that depends on the magnetization direction and is induced by the out-of-equilibrium current density ${\bf j}$ \cite{Hals2013}. In linear response, the torque is
\begin{align}\label{eq:generaltorque}
{\bs\tau}_{\alpha}={\bs m}_{\alpha}\times{\bs \eta}_{\alpha}\left[{\bs m}_{a},{\bs m}_{b}\right]{\bf j},
\end{align}
where the local second-rank tensor ${\bs \eta}_{\alpha}\equiv\left(\partial {\bf H}_{\alpha}/\partial {\bf j}\right)_{{\bf j}=0}$, encodes all forms of current-induced torques governed by the symmetries of ${\bs \eta}_{\alpha}$. Invoking Neumann's principle \cite{birss}, {\it symmetries possessed by the point group of the crystal will be inherited by any physical quantity}, we compute SOTs. 
In AFs, we consider symmetries that leaves the magnetic sub-lattice site invariant \cite{zelezny2017,internationalTableCryst}. Denoting ${\cal R}$ a symmetry rotation, the magnetization and current density obey, ${\bs m}'_{\alpha}=|{\cal R}|{\cal R}{\bs m}_{\alpha}$ and ${\bf j}'={\cal R}{\bf j}$ respectively, in the transformed frame. The transformation rule for the torque is ${\bs \tau}'_{\alpha}=|{\cal R}|{\cal R}{\bs \tau}_{\alpha}$. It follows from Neumann's principle that ${\bs\eta}_{\alpha}$ must satisfy, ${\bs\eta}_{\alpha}\left[{\bs m}'_{a},{\bs m}'_{b}\right]=|{\cal R}|{\cal R}{\bs\eta}_{\alpha}\left[{\bs m}_{a},{\bs m}_{b}\right]{\cal R}^T$. We consider the magnetic sublattices form inversion partners within each unit cell. This assumptions leads to the additional condition ${\bs\eta}_b[{\bs m}_b,{\bs m}_a]=-{\bs\eta}_a[{\bs m}_a,{\bs m}_b]$
%
\begin{figure}[ht]
\begin{center}
		\includegraphics[width=\columnwidth]{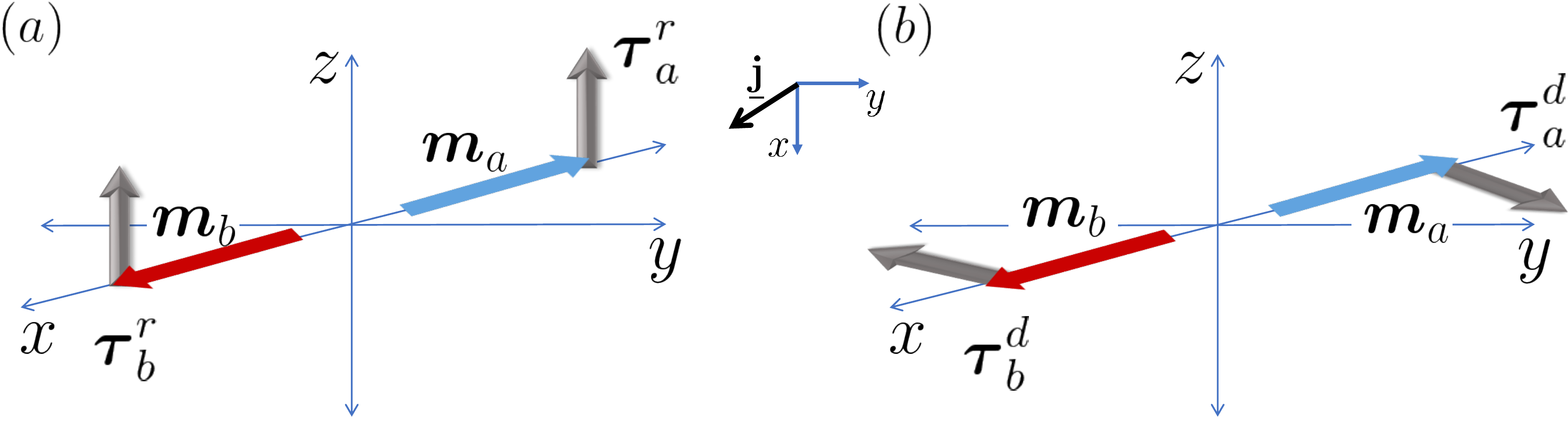}
		\caption{The action of (a) reactive ${\bs\tau}^r_{a,b}$ and (b) dissipative ${\bs\tau}^d_{a,b}$ spin-orbit torques on the magnetic moments at bipartite lattices, ${\bs m}_a$ and ${\bs m}_b$. The torques are driven by an in-plane current density $\underline{\bf j}$.}
		\label{fig:sot_display}
\end{center}
\end{figure}

To proceed, we simplify the tensor ${\bs\eta}_{\alpha}$ by expanding it up to leading order in ${\bs m}_{\alpha}$, i.e. ${\bs\eta}_{\alpha}={\bs\Lambda}_{\alpha}+{\bs\Lambda}^{\beta}_{\alpha}{\bs m}_{\beta}$, with $\beta=\{a,b\}$ \cite{Hals2013,birss}. The reactive and dissipative SOTs are captured by the second- and third-rank tensors, ${\bs\Lambda}_{\alpha}$ and ${\bs\Lambda}^{\beta}_{\alpha}$, respectively. The derived symmetry relations for ${\bs\eta}_{\alpha}$ imply that these tensors obey ${\bs\Lambda}_{a,b}=-|{\cal R}|{\cal R}{\bs\Lambda}_{b,a}{\cal R}^T$ and ${\bs\Lambda}^{a}_{a,b}=-{\cal R}{\cal R}{\bs\Lambda}^{b}_{b,a}{\cal R}^T$, for a specific symmetry group. 
	
We now focus on three groups of tetragonal non-centrosymmetric materials. First, we consider systems described by a linear in momentum Dresselhaus model \cite{zelezny2017,AdoPRB2017}. These are found in certain Heusler alloys, like the fully-compensated half-metallic ferrimagnet Mn$_2$Ru$_x$Ga (MRG) \cite{MRG2014,borisov2016tunnelling,vzic2016designing}. The large magnetocrystalline anisotropy and the lack of local inversion symmetry existing in MRG \cite{MRG2014} imply a considerable SOT. To compute the torques we consider that symmetries belongs to the tetragonal point group $\bar{4}2m$ \cite{MRG2014,internationalTableCryst}.  The simplest expression for SOTs is obtained when MRG is close to the cubic phase. As detailed in Supplemental Material \cite{SuppMat}, we find that the torques on each sublattice are
\begin{align*}
&{\bs\tau}_a=\Lambda{\bs m}_a\times \underline{\bf j}+{\cal T}
\left\{{\bs m}_a\times\left[\left(\Lambda^n{\bs m}_a+{\Lambda}^m{\bs m}_b\right)\times\underline{\bf j}\right]\right\},\\
&{\bs\tau}_b=-{\Lambda}{\bs m}_b\times \underline{\bf j}-
{\cal T}\left\{{\bs m}_b\times\left[\left({\Lambda}^m{\bs m}_a+{\Lambda}^n{\bs m}_b\right)\times\underline{\bf j}\right]\right\},
\end{align*}
for an in-plane charge current $\underline{\bf j}=(j_x,-j_y,0)$. The operator ${\cal T}\left\{\cdot \right\}$ switches the sign of $z$-component, while $x$- and $y$-components are unchanged. The phenomenological coefficients $\Lambda$, $\Lambda^n$ and $\Lambda^m$ quantify the strength of reactive and dissipative SOTs sketched in \figurename~\ref{fig:sot_display}. 
	
A second example are the AFs, Mn$_{2}$Au \cite{Mn2Au2018} and CuMnAs \cite{wadley2016}. The symmetries of the magnetic sites belongs to the polar point group $4mm$ \cite{zelezny2017}. The SOTs for these materials are similar to the torques in MRG in certain limits \cite{Mn2Au-CuMnAs-comment}. In fact, the result is obtained under the transformation $\underline{\bf j}\rightarrow{\bs z}\times{\bf j}$, with ${\bf j}=(j_x,j_y,0)$, and ${\cal T}\rightarrow I$, with $I$ the identity. Lastly, the third group of materials that we consider are part of enantiomorphic crystals \cite{zelezny2017,internationalTableCryst}. As an example we choose systems with point group $422$. The SOTs are found from the torques of MRG under the change $\underline{\bf j}\rightarrow{\bf j}$ and ${\cal T}\rightarrow I$. The expression found for these groups also recover the results given in Ref. \cite{zelezny2017}.  Details and the general result for the SOTs can be found in Supplemental Material \cite{SuppMat}.

{\it Antiferromagnetic dynamics.}- The magnetocrystalline anisotropy has two dominant components - an uniaxial anisotropy perpendicular to the AF layer and an  in-plane four-fold symmetric anisotropy. Next, we introduce the N\'eel vector ${\bs n}=\left({\bs m}_a-{\bs m}_b\right)/2$ and the magnetization ${\bs m}=\left({\bs m}_a+{\bs m}_b\right)/2$ that satisfy ${\bs m}\cdot{\bs n}=0$ and $m^2+n^2=1$. The free magnetic energy density, including the exchange energy is $f=-2\omega_{\text{ex}}{n}^2-\omega^{(1)}_{\parallel}n^2_z/2-\omega^{(1)}_{\perp}\left(n_x^4+n_y^4-6n^2_xn^2_y\right)/4-\omega^{(2)}_{\parallel}n^4_z$ in the macrospin description and employing the exchange approximation. The coupled equations of motion are
\begin{align}
\dot{\bs m}=&\label{eq:magEOM} {\bf f}_{m}\times{\bs m}+{\bf f}_{n}\times{\bs n}+\alpha\left({\bs m}\times\dot{\bs m}+{\bs n}\times\dot{\bs n}\right)+{\bs\tau}_m\\
\dot{\bs n}=&\label{eq:neelEOM} {\bf f}_{n}\times{\bs m} +{\bf f}_{m}\times{\bs n} +\alpha\left({\bs m}\times\dot{\bs n}+{\bs n}\times\dot{\bs m}\right)+{\bs\tau}_n
\end{align}
with the thermodynamic forces ${\bf f}_{m}\equiv-\partial f/\partial{\bs m}$ and ${\bf f}_{n}\equiv-\partial f/\partial{\bs n}$, and $\alpha$ the Gilbert damping constant. The SOTs exerted on the magnetization and N\'eel field are respectively defined as ${\bs\tau}_{m,n}=\left({\bs\tau}_a\pm{\bs\tau}_b\right)/2$.
In the exchange approximation, Eq. (\ref{eq:neelEOM}) implies that the small magnetization ${\bs m}$ is a slave variable obeying ${\bs m}=\left(\dot{\bs n}-{\bs\tau}_n\right)\times{\bs n}/(4\omega_{\text{ex}})$. Inserting the expression for ${\bs m}$ into Eq. (\ref{eq:magEOM}), we obtain the effective equation of motion that governs the N\'eel field dynamics,
\begin{align}\label{eq:effecNeel}
{\bs n}\times\left[\frac{\ddot{\bs n}}{4\omega_{\text{ex}}}-{\bf f}^{\text{an}}_{n}+\alpha_{\text{eff}}\left[{\bs n}\right]\dot{\bs n}+\Lambda\underline{\bf j}+{\bf X}\left[{\bs n},\dot{\bs n}\right]\right]=0,
\end{align}
with the constraint $|{\bs n}|=1$ and ${\bf f}^{\text{an}}_{n}$ the effective anisotropy field. We consider the simplest regime where $\Lambda^n$ and $\Lambda^m$ are comparable. ${\bf X}\left[{\bs n},\dot{\bs n}\right]$ is a field that derives from the dissipative SOT \cite{Xfield} and it is characteristic of MRG. However, when the easy-plane anisotropy $\omega^{(1)}_{\perp}$ is sufficiently large, this term can be neglected from Eq. (\ref{eq:effecNeel}). The dissipative torque also modifies the Gilbert damping by leading to the effective dissipation,
\begin{align}\label{eq:effectivedamping}
\alpha_{\text{eff}}\left[{\bs n}\right]=\alpha-\frac{\Lambda^{n}}{2\omega_{\text{ex}}}\underline{\bf j}\cdot{\bs n},
\end{align}
that depends on the charge current and the N\'eel vector.

For a comprehensive description of the AF dynamics, we restrict our analysis to the easy-plane magnetic configuration ($\omega^{(1)}_{\parallel}<0$, $\omega^{(2)}_{\parallel}\approx 0$ and $\omega^{(1)}_{\perp}>0$). In this case there are four in-plane degenerate states. We write ${\bs n}$ in terms of its polar ($\theta$) and azimuthal ($\phi$) angle. Maintaining the assumption of a dominant easy-plane anisotropy, we use the ansatz $\theta(t)=\pi/2$. Thus, Eq. (\ref{eq:effecNeel}) reduces to the scalar equation
\begin{align}\label{eq:phidynamics}
\frac{1}{4\omega_{\text{ex}}}\ddot{\phi}+\alpha_{\text{eff}}\left[\phi\right]\dot{\phi}=F[\phi],
\end{align}
that corresponds to a forced non-linear oscillator. The current controls the effective damping $\alpha_\text{eff}\left[\phi\right]$ of Eq. (\ref{eq:effectivedamping}) and the force $F[\phi]=-\partial V/\partial\phi$, where the potential
\begin{align}\label{eq:effpotential}
V[\phi]=-\frac{\omega^{(1)}_{\perp}}{4}\cos\left[4\phi\right]+\Lambda\left[j_x\cos\phi - j_y\sin\phi\right].
\end{align}
At equilibrium, when there is no current, the four degenerate ground states have azimuthal angles $\phi=-\pi$, $\phi=-\pi/2$, $\phi=0$ or $\phi=\pi/2$.


We will now demonstrate the emergence of self-oscillations. This is presented for MRG, however it is extensible to the other candidate materials under simple rotations of the charge current. We note that Eq. (\ref{eq:phidynamics}) describes a {\it Lienard system} \cite{perko} ensuring the existence and stability of steady-state oscillations. Furthermore, up to leading order in the amplitude of oscillations, Eq. (\ref{eq:phidynamics}) reduces to the well-studied Duffing-Van der Pol oscillator \cite{JENKINS2013,perko}. Our main result is that self-oscillations are stable in two scenarios. When the dissipative SOT $\Lambda^n$ is negative, self-oscillations emerge when the charge current is parallel to the $y$ or $x$ axis. In contrast, when $\Lambda^n>0$, the previous statement holds but stabilization occur in a narrow window of currents. Moreover, due to the four-fold symmetry, the azimuthal angle will behave the same for currents along $x$ or $y$. Therefore, the result for self-oscillations appearing when $\underline{\bf j}\parallel \hat{y}$, is replicated when $\underline{\bf j}$ is parallel to the $x$ axis by transforming $\phi\rightarrow\phi+\pi/2$. These results is detailed in the following.
	
\begin{figure}[ht]
	\begin{center}
		\includegraphics[width=\columnwidth]{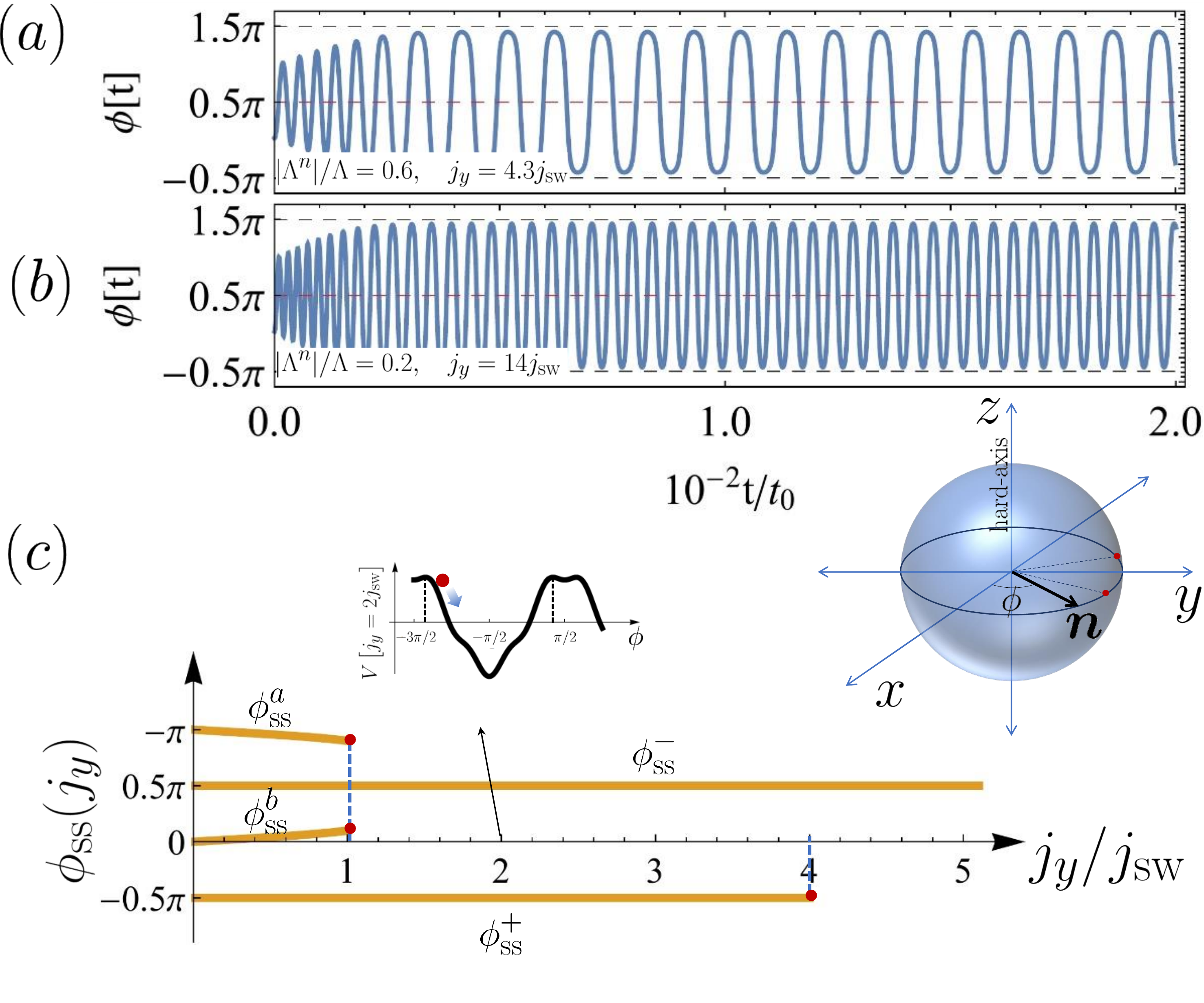}
		\caption{Time evolution of the azimuthal angle, in units of $t_0=1/\left(2\sqrt{\omega_{\text{ex}}\omega^{(1)}_{\parallel}}\right)$, driven by a current along the $y$-direction when the Gilbert damping is ${\alpha}=0.005$. The dissipative SOT strength and current are, (a) ${|\Lambda^n|}/{\Lambda}=0.6$ and $j_y=4.3j_{\text{sw}}$, (b) ${|\Lambda^n|}/{\Lambda}=0.2$ and $j_y=14j_{\text{sw}}$, respectively. (c) Steady-state angle as a function of current density $j_y$ normalized in $j_{\text{sw}}$. There is a global minima at $\phi^{-}_{\text{ss}}$ that prevails at any current. When $0\le j_y<4j_{\text{sw}}$, three steady-states exist (local minima) at $\phi^{a,b}_{\text{ss}}$ and $\phi^{+}_{\text{ss}}$, which become unstable at $j_y=j_{\text{sw}}$ and $j_y=4j_{\text{sw}}$, respectively.}
		\label{fig:solutions2}
	\end{center}
\end{figure}

We now consider a charge current along the $y$-axis. We first discuss the effects of the reactive torque, disregarding dynamical, and hence dissipative effects. The (reactive) current-dependent force $F_r=\Lambda j_y\cos\phi$ changes the azimuthal angle where the potential $V(\phi)$ attains its minimum. For positive current densities $0\le j_y<j_{\text{sw}}\equiv\omega^{(1)}_{\perp}/\Lambda$, there are four minima of $V(\phi)$, corresponding to the steady-states as seen in \figurename~\ref{fig:solutions2} (c). Among these, there are two degenerate configurations, $\phi^{a,b}_{\text{ss}}$, which become unstable when $j_y>j_{\text{sw}}$. The other two states are $\phi^{\pm}_{\text{ss}}$. However, when the current is sufficiently large, $j_y\ge 4 j_{\text{sw}}$, the state at $\phi^{+}_{\text{ss}}$ is unstable and thus leaving the only stable steady-state at $\phi^{-}_{\text{ss}}$, see Eq. (\ref{eq:effpotential}). For negative currents, $j_y < 0$, the steady-state angles reverse their sign, $\phi_{\text{ss}} (j_y)=-\phi_{\text{ss}} (-j_y )$.
	
We now include the dissipative SOT and show how self-oscillations emerge from Eq. (\ref{eq:phidynamics}).  The essential requirement for self-oscillations is that the dissipative spin-orbit forces compensate the magnetization dissipation. To find the onset of self-oscillations, it is sufficient to evaluate the effective damping Eq. (\ref{eq:effectivedamping}) around the steady-state $\phi^{-}_{\text{ss}}$. The damping is then ${\alpha}^{y}_{\text{eff}}={\alpha} + \left(\Lambda^{n}/2\omega_{\text{ex}}\right) j_y\sin\left[\phi_{\text{ss}}(j_y)\right]$, which is non-linear in $j_y$ because  $\phi_{\text{ss}}$ depends on the current. Consequently, in order to compensate the Gilbert damping, the dissipative strength has to be negative since $\phi^{-}_{\text{ss}} > 0$ \cite{jy_lambdaApos}. This result is valid when the current flows in both directions $\pm\hat{y}$, since the steady-state angle is an odd function of the current. Self-oscillations set in when the current exceeds the threshold $j^{(1)}_{\text{so}}=2\omega_{\text{ex}}\alpha/|\Lambda^{n}|$, where the damping satisfy $\alpha^{y}_{\text{eff}}=0$.
	
We numerically solve Eq. (\ref{eq:phidynamics}) for the azimuthal angle $\phi(t)$ in two representative cases. In these simulations, the Gilbert damping constant is $\alpha=0.005$ and there is a small anisotropy field $\omega^{(1)}_{\perp}/\omega_{\text{ex}}=10^{-2}$ consistent with the exchange approximation. The ratio  of the dissipative and reactive SOTs strength is considered in the range ${|\Lambda^n|}/{\Lambda}\in \left[0.1,1\right]$. \figurename~\ref{fig:solutions2}(a) shows the time-evolution of $\phi(t)$ when there is a charge current $j_y=4.3 j_{\text{sw}}$ and ${|\Lambda^n|}/{\Lambda}=0.6$. As an initial condition, we slightly perturb the steady-state $\phi^{-}_{\text{ss}}$ so that the azimuthal angle starts at  $\phi(t=0)=\phi^{-}_{\text{ss}}+0.01$. Thereafter, the azimuthal angle oscillates around $\phi^{-}_{\text{ss}}$ with an amplitude that increases until it becomes stable. We find that the saturated amplitude is independent on the initial conditions, as long as $\phi(t=0)\neq\phi^{-}_{\text{ss}}$. In a second example, displayed in \figurename~\ref{fig:solutions2}(b), we consider the evolution of $\phi(t)$, with the same initial condition as above, but ${|\Lambda^n|}/{\Lambda}=0.2$ and under a larger current $j_y=14j_{\text{sw}}$. As we see, $\phi(t)$ stabilizes at an amplitude similar to the previous example but twice of frequency.

The numerical results can be understood intuitively. In order to prevent N\'eel-field oscillations from relaxing into the current-induced steady-state determined by the reactive torque, a significant dissipative SOT is paramount. Realizing stable oscillations require a large charge current when the SOT is small. When $j_y\ge j^{(1)}_{\text{so}}$, and after the transient features, the terminal amplitude of the oscillations is a good order parameter to characterize the stable regimes. We compute pertubatively the equation of motion for the amplitude of oscillations $\cal{A}$ using {\it multiple-scale analysis} \cite{StrogatzBook}. Two natural time scales dominate the evolution of $\phi(t)$, see \figurename~\ref{fig:solutions2}(a). There are rapid precessions around $\phi^{-}_{\text{ss}}$. Additionally, the amplitude of precessions evolves on a slower time scale. Averaging over the rapid precessions, to leading order in the amplitude ${\cal A}$, we find that the amplitude obeys the well-known Ginzburg-Landau equation \cite{perko}, 
\begin{equation}
2\dot{\cal{A}}={\cal{A}}-{\cal{A}}^{3}/\left[8\left(1-j^{(1)}_{\text{so}}/j_y\right)\right].
\label{GL}
\end{equation} 
At long time scales, Eq.\ (\ref{GL}) determines the terminal amplitude {$\bar{\cal{A}}=2\sqrt{2}\sqrt{1-j^{(1)}_{\text{so}}/j_y}$}. Consequently, the amplitude $\bar{\cal{A}}\rightarrow 0$ when $ j_y \le j^{(1)}_{\text{so}}$, reproducing the boundary between static states and self-oscillations. For large current densities $j_y\gg j^{(1)}_{\text{so}}$, the amplitude saturates to $\bar{\cal{A}}\approx2\sqrt{2}$. This analytical result is somewhat smaller than the value reached by the azimuthal angle in the numerical solution shown in Fig. \ref{fig:solutions2}. This is because an exact description of large-amplitude oscillations requires the inclusion of higher order terms in the analytical treatment of $\cal{A}$, which is beyond the scope of the present study.

\begin{figure}[h!]
	\begin{center}
		\includegraphics[width=\columnwidth]{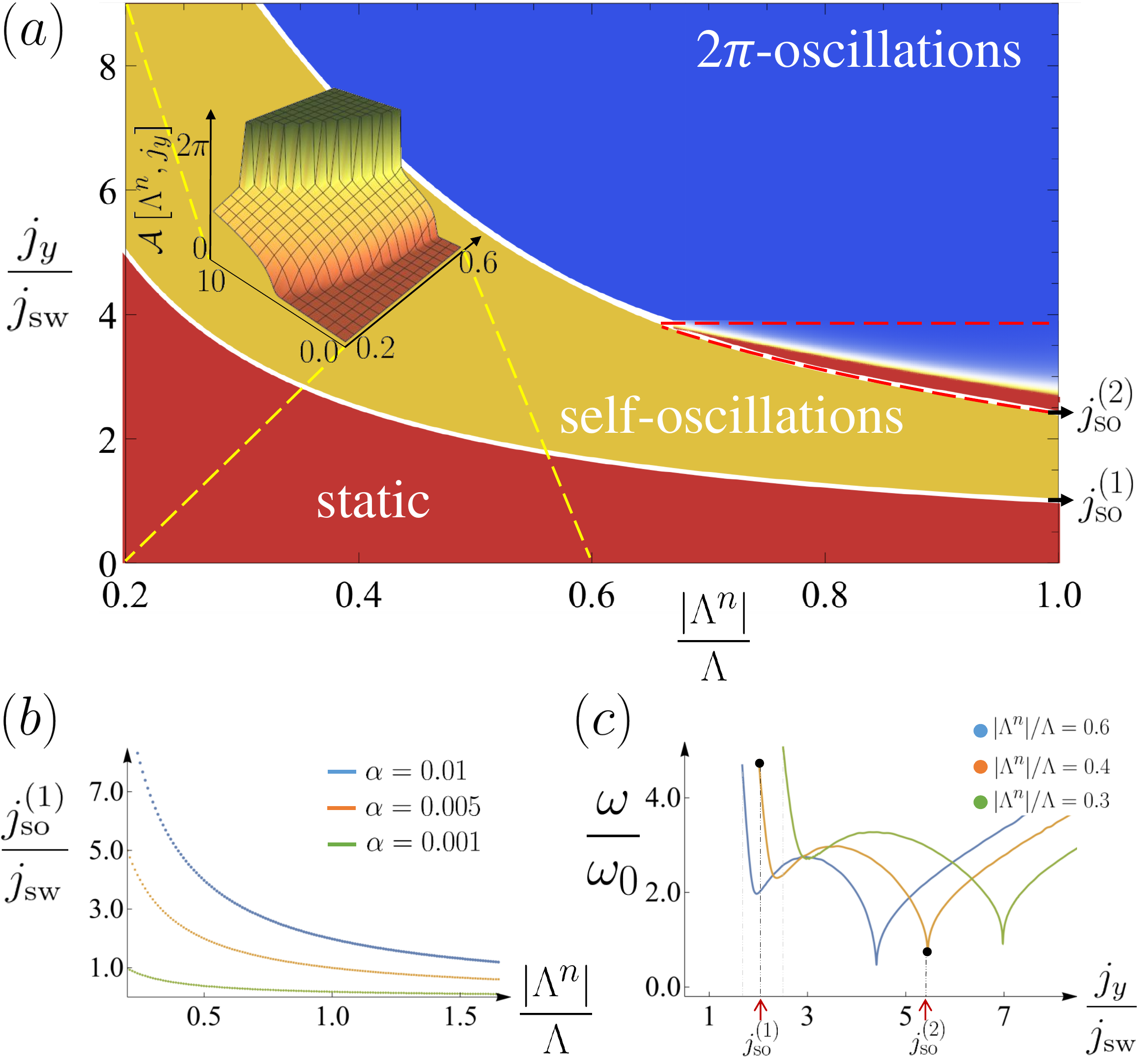}
		\caption{(a) Terminal amplitude of self-oscillations as a function of current density $j_y$ and dissipative SOT strength $|\Lambda^n|$. (b) The threshold current $j^{(1)}_{\text{so}}$, defining the border between static and stable oscillations, for different values of ${\alpha}$. For $\alpha=0.005$, in (c) is shown in units of $\omega_0=1/t_0$ the characteristic frequency of self-oscillations in terms of the current density. At the threshold $j^{(1)}_{\text{so}}$, the frequencies drop to zero corresponding to the static regime. At $j^{(2)}_{\text{so}}$ the frequency reaches a finite value that change in terms of the ratio $|\Lambda^n|/\Lambda$. For $j_y > j^{(2)}_{\text{so}}$, the frequency of full oscillations grows strictly with the current.}
		\label{fig:phasediagram_jy}
	\end{center}
\end{figure}
	
Additionally, we numerically evaluate the terminal amplitude ${\cal{A}}\left[\Lambda^n,j_y\right]$, for various charge currents $j_y$ and dissipative torques $\Lambda^n$.  The phase diagram (\figurename~\ref{fig:phasediagram_jy}(a)) shows three regions; static, self-oscillations, and $2\pi$ self oscillations. The first two are separated by the threshold current $j^{(1)}_{\text{so}}$ shown in \figurename~\ref{fig:phasediagram_jy}(b) for different values of damping. When ${|\Lambda^n|}/{\Lambda}$ is sufficiently large, the N\'eel field dynamics either vanish or proliferate into complete $2\pi$ precessions. As the dissipative strength decrease, an intermediate window of stable (amplitude $< 2 \pi$) oscillations open up: the lesser $|\Lambda^n|$ is, the wider the window is. In this region, the amplitude grows with increasing dissipative SOT, i.e., increasing $|\Lambda^n|$ and/or current density $j_y$. Eventually, when $j_y$ is sufficiently large, the oscillator enters the proliferation phase. The threshold current $j^{(2)}_{\text{so}}$ separates ${\cal A } < 2\pi$ from ${\cal A } = 2\pi$. If this threshold satisfy $j^{(2)}_{\text{so}}<4j_{\text{sw}}$, the terminal amplitude has two possible outcomes ${\cal A}=0$ or ${\cal A}=2\pi$. This region in the phase diagram is delimited by a red dashed line in \figurename~\ref{fig:phasediagram_jy}(a). The colormap therein represents the probability \cite{prob} to find: static solutions (red region) for currents close to $j^{(2)}_{\text{so}}$ and $2\pi$-amplitude oscillations (blue region) when the current is near to $4j_{\text{sw}}$. This indeterminate behaviour is due to local minima in $V[\phi]$  appearing at $\phi^{+}_{\text{ss}}=-\pi/2$ when $j_{\text{sw}}<j_y<4j_{\text{sw}}$, see inset in \figurename~\ref{fig:solutions2}(c). When the amplitude of self-oscillations reaches its maximum value, small increases on the current or small perturbations in the initial conditions can make the evolution of $\phi(t)$ to decay into $\phi^{+}_{\text{ss}}$ or proliferate in complete oscillations. 
	
In the region $j^{(1)}_{\text{so}} < j_y < j^{(2)}_{\text{so}}$, the oscillator frequency has a non-trivial dependence on the current, as shown in \figurename~\ref{fig:phasediagram_jy}(c). In contrast to the $2\pi$-amplitude oscillating phase, the frequency of the ${\cal A}<2\pi$-phase drops to a finite (non-zero) value at $j^{(2)}_{\text{so}}$. Also, a second, deeper drop occurs close to $j^{(1)}_{\text{so}}$ and it is deeper as the ratio $|\Lambda^n|/\Lambda$ decreases. When full oscillations occur, marked by $j_y\ge j^{(2)}_{\text{so}}$, the frequency monotonically grows and eventually becomes linear for large currents.
	
In the self-oscillating phase, a dc current controls the ac output power. This is because the N\'eel vector precesses around the direction of the applied dc current, either along the $x$- or the $y$-direction. Through the anisotropic magnetoresistance (AMR) \cite{Mn2Au2018,AMR_AF_sem}, the relative change of the N\'eel vector with respect to the direction of current changes the induced voltage across the device. This transforms the self-oscillations into an oscillating voltage signal. 
	
We predict that the self-oscillating phase appears for currents in the order of critical currents required for switching the magnetic moment. Measurements of the reactive and antidamping SOTs, predicted using symmetry arguments, will help to pave the way for self-oscillations in AFs. Recent measurements on the compensated ferrimagnet Mn$_2$Ru$_x$Ga \cite{MRG2014,borisov2016tunnelling,vzic2016designing} indicate a considerable dissipative SOT that enables the realization of our proposal. While current-induced magnetization dynamics studies in Mn$_2$Ru$_x$Ga are unexplored, we expect this material to exhibit a self-oscillating regime in the Terahertz gap due to its low Gilbert damping, high anisotropy and strong spin-orbit coupling. An additional advantage of this material is the possibility to tune the sign of the $\Lambda^n/\Lambda$ ratio by the compensation temperature as well as the prospect to use the stronger magneto-resistive effects associated with highly spin polarised materials to convert the dc charge current into an oscillating ac voltage.

{\it Conclusion and discussion.}-- In conclusion, we have demonstrated that electrical currents can control antiferromagnetic self-oscillations. Using symmetry arguments on tetragonal crystalline antiferromagnets, we establish that SOTs contain reactive contributions and essential dissipative terms. By expressing the SOTs in terms of phenomenological parameters, we demonstrate that a non-linear driven damped oscillator describes the antiferromagnetic dynamics. The dissipative SOT controls the effective damping enabling oscillations. When the charge current exceeds a threshold, self-oscillations spontaneously develop. An important feature of the self-oscillating phase is that the current density controls the ac output power. The N\'eel vector precesses around the direction of current. In turn, this generates an oscillating voltage signal via the anisotropic magnetoresistance effect (AMR) \cite{Mn2Au2018}.
	
\begin{acknowledgments}
RT would like to thank Alejandro Leon for disussions. This work was supported by the European Union's Horizon 2020 Research and Innovation Programme under Grant DLV-737038 "TRANSPIRE" and the Research Council of Norway through is Centres of Excellence funding scheme, Project No. 262633, "QuSpin". The European Research Council via Advanced Grant number 669442 "Insulatronics".

Note added--  During the completion of our work, we became aware of another paper \cite{AutoOsPy2018} that studied auto-oscillations in single permalloy layers using spin-orbit torques.
\end{acknowledgments}

\bibliography{sot-so}

\begin{thebibliography}{36}
\bibitem [{jy_()}]{jy_lambdaApos}%
  \BibitemOpen
  \href@noop {} {\bibinfo  {journal} {Self-oscillations around
  $\phi^+_{\text{ss}}$ are obtained when $\Lambda^n>0$. These oscillations are
  highly unstable and thus, are not considered}\ }\BibitemShut {NoStop}%
\end{thebibliography}

\end{document}